\documentclass[aps, prd, superscriptaddress, twocolumn]{revtex4-1}

\usepackage{wallpaper}
\usepackage{bm}
\usepackage{amsmath}
\usepackage{amssymb}
\usepackage[colorlinks,linkcolor=blue,anchorcolor=blue,citecolor=red]{hyperref}

\usepackage{booktabs}
\usepackage{array}
\usepackage{graphicx}

\begin{document}

\title{Magnetized Coulomb crystals in neutron star crusts}

\author{Cheng-Jun~Xia}
\email{cjxia@yzu.edu.cn}
\affiliation{Center for Gravitation and Cosmology, College of Physical Science and Technology, Yangzhou University, Yangzhou 225009, China}

\author{Toshiki Maruyama}
\email{maruyama.toshiki@jaea.go.jp}
\affiliation{Advanced Science Research Center, Japan Atomic Energy Agency, Shirakata 2-4, Tokai, Ibaraki 319-1195, Japan}

\author{Nobutoshi Yasutake}
\email{nobutoshi.yasutake@it-chiba.ac.jp}
\affiliation{Department of Physics, Chiba Institute of Technology (CIT), 2-1-1 Shibazono, Narashino, Chiba, 275-0023, Japan}
\affiliation{Advanced Science Research Center, Japan Atomic Energy Agency, Shirakata 2-4, Tokai, Ibaraki 319-1195, Japan}

\author{Toshitaka Tatsumi}
\email{tatsumitoshitaka@gmail.com}

\date{\today}

\begin{abstract}
We investigate the properties of magnetized Coulomb crystals in neutron star crusts using a fully three-dimensional geometry with periodic boundary conditions. The electron density profiles are fixed via the Thomas-Fermi approximation, and the nonuniform magnetic fields are treated with the equivalent magnetic charge method. The study focuses on Coulomb crystals composed of $^{12}$C at an ion number density $n_d = 10^{-9}\,\text{fm}^{-3}$, subjected to various external magnetic fields. Nuclei are described by a Gaussian wave function, where the width $b$ encapsulates the effects of zero-point ion vibrations and finite temperature. Our findings show that the crystal softens as $b$ increases. The Madelung constant $K_M$ fluctuates with the external magnetic field $B_{z0}$ at $B_{z0}\leq 3\times 10^{14}$\,G. At higher field strengths, $K_M$ increases until $B_{z0} \approx 3\times 10^{15}$\,G and then decreases. The body-centered cubic (BCC) lattice is slightly more stable than the face-centered cubic (FCC) lattice when $B_{z0} < 3\times 10^{15}$\,G, whereas the FCC lattice may become more stable at larger $B_{z0}$. The elastic constants $c_{11}-c_{12}$ and $c_{44}$ are computed and tabulated, which grow with $B_{z0}$ for $3\times 10^{14}\,\mathrm{G}\lesssim B_{z0}\lesssim 2\times 10^{15}$\,G and then decline toward zero as the field strength increases further. For $B_{z0}\gtrsim 10^{16}$\,G, it becomes difficult to identify a stable lattice structure. These results provide valuable insights into the role of strong magnetic fields in shaping the properties of Coulomb crystals in compact stars.
\end{abstract}


\maketitle

\section{\label{sec:intro}Introduction}
According to stellar evolution models, more than 95\% stars will end up as white dwarfs, which are typically made of $^{12}$C and $^{16}$O with the density reaching $\sim$10$^6$ g/cm$^3$ in their cores~\cite{Saumon2022_PR988-1}. Similar structures are also expected in neutron star crusts~\cite{Feynman1949_PR75-1561, Baym1971_ApJ170-299}. As compact stars slowly cool down, the Coulomb interactions among nuclei become important, which can be characterized by the ratio of Coulomb energy to thermal energy for one-component plasma, i.e., the Coulomb parameter:
\begin{equation}
  \Gamma=\frac{e^2Z^2}{R_\mathrm{W} T}. \label{Eq:Gamma}
\end{equation}
Here $Z$ is the proton number of nuclei,  $T$ the temperature, and $R_\mathrm{W}  = \left(4\pi n_d/3\right)^{-1/3}$ the Wigner-Seitz (WS) cell size with $n_d$ being the nuclei density and the subscript $d$ represents nuclei (droplets). The coefficient $e^2$ ($=1/137.03599976$) corresponds to the dimensionless fine structural constant since in this work we employ natural units with $\hbar = c = 1$.  For small values of $\Gamma$ ($\ll 1$), Coulomb interactions are insignificant and ions behave like an ideal noninteracting gas. As $\Gamma$ increases to $\sim$1, Coulomb interactions become important, where ions undergo short-range correlations and behaving like a liquid~\cite{Althaus2010_AAR18-471}. For larger $\Gamma$ surpassing some critical value, e.g., the melting parameter $\Gamma_m\approx175$ fixed by Monte Carlo simulations~\footnote{The exact criterion may be altered by electron screening~\cite{Potekhin2000_PRE62-8554}.},  Coulomb interactions dominate over thermal fluctuations. In such cases, ions will eventually freeze and form a crystallized structure~\cite{Brush1966_JCP45-2102, Potekhin2000_PRE62-8554}, which corresponds to a first-order liquid-solid phase transition with the transition temperature determined by fulfilling $\Gamma = \Gamma_m$.

The energy density of a cold Coulomb crystal is determined by~\cite{Chamel2020_PRC101-032801}
\begin{equation}
  E = m_d n_d + \frac{2 \pi + e^2}{2 \pi} E_e + K_M \sigma(Z) \mu_{0}, \label{eq:Et}
\end{equation}
with $m_d$ being the mass of nuclei and $\mu_{0}\equiv n_d e^2 Z^2/{R_\mathrm{W}}$. The average electron number density is fixed by $n_e=Z n_d$, which determines the energy density of free electron gas $E_e$. The last term in Eq.~(\ref{eq:Et}) corresponds to the lattice energy density with electron polarization corrections. For a body-centered cubic (BCC) lattice, the Madelung constant $K_M = -0.895929255682$~\cite{Baiko2001_PRE64-57402} and the polarization factor $\sigma(Z)$~\cite{Chamel2020_PRC101-032801, Potekhin2000_PRE62-8554} is given by
\begin{equation}
\sigma(Z) = 1 + \frac{ 12^{{4}/{3}}  Z^{{2}/{3}}e^2}{35 \pi^{{1}/{3}}} \left(1-\frac{ 1.1866}{Z^{ 0.267}}+\frac{ 0.27}{Z}\right). \label{eq:plrz}
\end{equation}

The elastic properties of Coulomb crystals were extensively investigated in previous works~\cite{Fuchs1936_PRSA157-444, Ogata1990_PRA42-4867, Strohmayer1991_ApJ375-679}. Assuming point nuclei embedded in a uniform electron background, the elastic constants of BCC and face-centered cubic (FCC) crystals at vanishing temperatures are obtained with~\cite{Ogata1990_PRA42-4867}
\begin{equation}
 \left\{\begin{array}{l}
   \mathrm{BCC:}\ c_{11}-c_{12}=0.04908 \mu_{0}, \ c_{44}=0.1827 \mu_{0}; \\
   \mathrm{FCC:}\ c_{11}-c_{12}=0.04132 \mu_{0}, \ c_{44}=0.1852 \mu_{0}. \\
 \end{array}\right. \label{Eq:El_MC}
\end{equation}
Due to the zero-point ion vibrations at lower temperatures, the shear modulus may be reduced by up to 18\% for light nuclei~\cite{Baiko2011_MNRAS416-22}, while the modification becomes negligible for larger nuclei. Meanwhile, it was shown that the effects of electron screening in the randomly oriented polycrystalline matter also lead to a reduction on the shear modulus~\cite{Kobyakov2013_PRC87-055803, Kobyakov2015_MNRAS449-L110}. The elastic properties of multicomponent crystals were examined as well, which agrees with the results obtained from the linear mixing rule~\cite{Kozhberov2019_MNRAS486-4473, Chugunov2020_MNRAS500-L17}.

It is well known that there exist strong magnetic fields in neutron stars, which could reach $10^{14\text{-}15}$ G on magnetars' surfaces and even $\sim$$10^{18}$ G inside~\cite{Cardall2001_ApJ554-322}. Under the strong magnetic fields, the properties of Coulomb crystals are expected to be altered~\cite{Lai2001_RMP73-629, Baiko2017_MNRAS470-517, Roy2019_PRD100-063008}. In particular, for a stationary system with magnetic field $\bm{B}(\bm{r})$, the momentum conservation equation~\cite[Eq.~(8)]{Carroll1986_ApJ305-767} is reduced into
\begin{equation}
  \sum_{i} \frac{\partial}{\partial r_i} \left(\bar{\sigma}_{i j} +  \frac{2 B_i B_j - B^2 \delta_{i j} }{8\pi} \right)   = 0. \label{Eq:hydroeq}
\end{equation}
Here the indices $i$ and $j$ correspond to the Cartesian components ($x$, $y$, and $z$), while the stress tensor $\bar{\sigma}_{i j}$ is related to the variation of energy density with respect to a (deformed) Coulomb crystal $\delta E = \sum_{i,j} \bar{\sigma}_{ij}u_{ij}$ with $u_{ij}$ being a displacement gradient. Evidently, the Coulomb crystals will be stretched and/or squeezed at certain directions in the presence of a non-uniform magnetic field~\cite{Baiko2017_MNRAS470-517}, which may be produced by magnetohydrodynamic evolutions of a non-stationary liquid system as it freezes or magnetic field evolution in the solid phase.

On the microscopic scale, the non-uniform magnetic fields may also arise from the magnetization of Coulomb crystals, which could in turn affect their lattice structures and elastic properties. In this work, employing a fully three-dimensional geometry with periodic boundary condition~\cite{Okamoto2012_PLB713-284, Okamoto2013_PRC88-025801, Xia2021_PRC103-055812, Xia2022_PRD106-063020, Xia2023_PLB839-137769}, we explore the properties of magnetized dense stellar matter in neutron star crusts. The Thomas-Fermi approximation (TFA) is employed to fix the electron density profiles, while the nonuniform magnetic fields are treated with equivalent magnetic charge method.

The paper is organized as follows. In Sec.~\ref{sec:the} we present the theoretical framework for obtaining the properties of magnetized Coulomb crystals. Adopting various external magnetic fields, the Madelung constants and elastic properties of BCC and FCC lattices composed of $^{12}$C at an ion number density $n_d = 10^{-9}\,\text{fm}^{-3}$ are examined in Sec.~\ref{sec:res}. We draw our conclusion in Sec.~\ref{sec:con}.

\section{\label{sec:the} Theoretical framework}

\subsection{\label{sec:the_Crystal} Coulomb crystals}
\subsubsection{\label{sec:the_Clm} Electromagnetic fields}
The electromagnetic field can be obtained by solving
\begin{equation}
\nabla^2 A^\mu = -\sum_i q_i J_i^\mu, \label{eq:KG_photon_mean}
\end{equation}
with the currents $J_i^\mu = \langle \bar{\Psi}_i\gamma^\mu \Psi_i \rangle$. If we neglect the net electric currents within the system,
Eq.~(\ref{eq:KG_photon_mean}) can be simplified and gives
\begin{eqnarray}
\nabla^2 A_0 &=& -e\rho_\mathrm{ch} =e\rho_e -  \sum_i q_i f(\bm{r}-\bm{r}_i), \label{eq:KG_A0} \\
\nabla^2 \Phi_M &=& \nabla\cdot\mathbf{M} = - \rho_M. \label{eq:KG_mag}
\end{eqnarray}
Here $f(\bm{r}) = \phi(\bm{r})^2$ the nuclei distribution function, which is determined by the wave function of a nucleus assuming gaussian form, i.e.,
\begin{equation}
  \phi(\bm{r}) = \left( \frac{1}{\pi b^2} \right)^{\frac{3}{4}} \exp\left(-\frac{\bm{r}^{2}}{2b^2} \right). \label{eq:Gaus}
\end{equation}
The coefficient $b=1/\sqrt{m_d \omega_0}$, corresponding to the lowest energy level $3\omega_0/2$ for a harmonic oscillator in a potential
\begin{equation}
  V(\bm{r})=\frac{1}{2}m_d \omega_0^2 \bm{r}^2, \label{eq:VHO}
\end{equation}
so that the zero-point ion vibrations may be effectively considered. The delocalizing effects of nuclei arise from finite temperature may also be considered effectively by further increasing $b$. The magnetization $\mathbf{M}$ in Eq.~(\ref{eq:KG_mag}) is obtained with
\begin{equation}
  \mathbf{M} = \mathbf{M}_e+\sum_i \mathbf{m}_i f(\bm{r}-\bm{r}_i),
\end{equation}
where $\mathbf{M}_e$ represents the magnetization of electrons, $\mathbf{m}_i$ is the magnetic moment of nucleus $i$. Then the effective magnetic charge density $\rho_M$ and magnetic scalar potential $\Phi_M$ can be obtained with Eq.~(\ref{eq:KG_mag}), which determines the magnetic field with
\begin{equation}
  \mathbf{B} = \mathbf{M} + \mathbf{H} = \mathbf{M} - \nabla \Phi_M + \mathbf{H}_0. \label{eq:Bcal0}
\end{equation}
Here $\mathbf{H}_0 = \mathbf{B}_0$ is adopted to introduce an external homogeneous magnetic field. The magnetization $\mathbf{M}_e$ can be obtained with~\cite{LANDAU1984105, Broderick2000_ApJ537-351, Rabhi2015_PRC91-045803}
\begin{equation}
  \mathbf{M}_e = - \left.\frac{\mbox{d} \mathcal{E}_e}{\mbox{d} B}\right|_{\rho_e} \frac{\mathbf{B}}{B} =  \mathcal{M}_e \frac{\mathbf{B}}{B} =  \chi_e \mathbf{B},
\end{equation}
where $\mathcal{M}_e$ and $\chi_e$ represent the magnetization and magnetic susceptibility of electrons.
Then Eq.~(\ref{eq:Bcal0}) can be simplified and gives
\begin{equation}
  \mathbf{B} = \frac{\mathbf{H}_0 - \nabla \Phi_M} {1- \chi_e - \chi_I}. \label{eq:Bcal}
\end{equation}
Here $\chi_I = \sum_i \mathbf{m}_i f(\textbf{r}-\textbf{r}_i)/{\mathbf{B}}$ is the nuclear magnetic susceptibility. Note that $\rho_e$,  $\chi_e$, $\chi_I$ and $\mathcal{E}_e$ represent the local properties of electron gas and vary with the space coordinate $\bm{r}$ in TFA, which can be determined by fulfilling the constancy of the chemical potential $\mu_e(\bm{r})$ at given $A_0$ and $\mathbf{B}$.

Based on the mean field approximation, the total energy of the system is obtained with
\begin{eqnarray}
E &=& \int \left[\mathcal{E}_e + \frac{1}{2} e A_0 \rho_\mathrm{ch} + \frac{1}{2} B^2 - \sum_i  \mathbf{m}_i \cdot \mathbf{B} f(\bm{r}-\bm{r}_i)  \right] \mbox{d}^3 r \nonumber \\
 && +\sum_i(m_i- \Sigma_i),  \label{eq:energy}
\end{eqnarray}
where $m_i$ is the mass and $\Sigma_i = \int \left[\frac{1}{2} (\nabla A_{0,i})^2 + \frac{1}{2} B_i^2  \right] \mbox{d}^3 r$ the self electromagnetic energy of the nucleus with $A_{0,i}$ and $B_i$ being the corresponding electromagnetic fields determined by $q_i$, $\mathbf{m}_i$, and $\phi(\bm{r})$.

\subsubsection{Local properties \label{sec:ch}}
At zero temperature, the local number density and energy density of electrons in TFA read
\begin{eqnarray}
  \rho_e(\bm{r}) &=& \sum_s \int \Theta[\mu_e(\bm{r}) - \epsilon_e(\bm{r})] \frac{\mbox{d}^3 p}{(2\pi)^3}, \label{eq:rhof_0}\\
  \mathcal{E}_e(\bm{r}) &=& \sum_s \int \Theta[\mu_e(\bm{r}) - \epsilon_e(\bm{r})] \sqrt{p^2+{m_e}^2} \frac{\mbox{d}^3 p}{(2\pi)^3}.  \label{eq:ef_0}
\end{eqnarray}
Here $\mu_e(\bm{r})$ and $\epsilon_e(\bm{r})$ represent the chemical potential and single particle energy of electrons. The momentum of charged spin one-half electrons perpendicular to magnetic field $\textbf{B}$ is quantized with
\begin{equation}
 p_\bot^2 = 2 n e B,
\end{equation}
where the quantum number $n$ enumerates the Landau levels and is linked to the spin $s=\pm 1$ and orbital angular momentum
$l=0, 1,2,\ldots$ via
\begin{equation}
n=l + \frac{1}{2} + \frac{s}{2}.
\end{equation}
The corresponding single particle energies are
\begin{equation}
\epsilon_e(p_\|, n) =  \sqrt{p_\|^2+{\bar{m}_e}^2} -e A_0,
\label{eq:eigen}
\end{equation}
where
\begin{equation}
\bar{m}_e(n) = \sqrt{ {m_e}^2 + 2 n e B}.
\end{equation}
The density of states is modified via
\begin{equation}
2\int \frac{\mbox{d}^3 p}{(2\pi)^3} \rightarrow \frac{eB}{2\pi^2} \sum_{s=\pm 1}\sum_{l=0}^{n(l,s)\leq n_e^\mathrm{max}}
                \int_0^{\nu_e(l,s)} \mbox{d} p_\|,
\end{equation}
with
\begin{eqnarray}
n_e^\mathrm{max} &=& \mathrm{int}\left[\frac{(E^f_e)^2 - {m_e}^2}{2 e B}\right], \\
\nu_e(n) &=& \sqrt{(E^f_e)^2 - {\bar{m}_e}^2}.
\end{eqnarray}
Here the Fermi energy $E^f_e = \mu_e + e  A_0$ is determined by the local chemical potential $\mu_e$, and $\nu_e(n)$ represents the Fermi momentum of electrons corresponding to the principal quantum number $n$. The local electron number density, energy density, and magnetization can be obtained with~\cite{Broderick2000_ApJ537-351, Huang2010_PRD81-045015, Strickland2012_PRD86-125032, Dong2013_PRD87-103010, Rabhi2015_PRC91-045803}
\begin{eqnarray}
\rho_e &=& \frac{eB}{2\pi^2} \sum_{l,s}^{n\leq n_e^\mathrm{max}} \nu_e, \label{eq:rhoi_c}\\
\mathcal{E}_e &=& \frac{eB}{4\pi^2} \sum_{l,s}^{n\leq n_e^\mathrm{max}}
               \left[\nu_e E^f_e + \bar{m}_e^2 \ln{\left|\frac{\nu_e + E^f_e }{\bar{m}_e}\right|}\right],  \label{eq:ei_c}\\
\mathcal{M}_e &=& \frac{E^f_e\rho_e - \mathcal{E}_e}{B} - \frac{e^2 B}{2\pi^2} \sum_{l,s}^{n\leq n_e^\mathrm{max}} n \ln{\left|\frac{\nu_e + E^f_e }{\bar{m}_e}\right|}. \label{eq:Mi_c}
\end{eqnarray}
Note that the Larmor radius is approximately $\lambda_e B_q/B$ with $\lambda_e = 1/m_e = 2426$ fm and $B_q=m_e^2/e=4.414\times 10^{13}$ G, which validates the TFA adopted in this work since it is much smaller than the size of the system.

\subsection{\label{sec:the_elestic} Elasticity theory}
In the framework of elasticity theory, the variation of energy density due to deformation is determined by~\cite{Wallace1967_PR162-776}
\begin{equation}
\delta E = \sum_{i,j} \sigma_{ij}u_{ij} + \sum_{i,j,k,l}\frac{1}{2}S_{ijkl}u_{ij}u_{kl}. \label{eq:dE0}
\end{equation}
Here the indices $i$, $j$, $k$, $l$ correspond to the Cartesian components ($x$, $y$, and $z$), $\sigma_{ij}$ ($=-P\delta_{ij}$) the stress tensor in an undeformed solid, and $S_{ijkl}$ the elastic modulus tensor. The deformation is introduced via a displacement gradient $u_{ij}$, where an ion at position $\vec{r}$ is moved to a new position $\vec{r}'$ with
\begin{equation}
  r'_i = r_i + \sum_j u_{ij} r_j.
\end{equation}
Here we take $u_{ij}$ as constants, corresponding to a uniform deformation. If the Coulomb crystal deforms at a fixed volume, i.e., $\mathrm{det}\left[u+I_{3\times3}\right]=1$, the first term in {Eq.~(\ref{eq:dE0})} vanishes. Carrying out a transformation ($ij$, $kl$) $\rightarrow$ ($m$, $n$) of the subscripts ($xx$, $yy$, $zz$, $xy$, $yz$, $zx$)  $\rightarrow$ (1, 2, 3, 4, 5, 6), the elastic constants $c_{mn}$ are connected to the elastic modulus tensor with $c_{mn} = S_{ijkl}$~\cite{Ogata1990_PRA42-4867}.
Then the variation of energy density becomes
\begin{equation}
\delta E = \frac{1}{2} \sum_{n=1}^3\sum_{m=1}^3 c_{mn} u_m u_n + 2\sum_{m=4}^6 c_{mm} u_m^2, \label{eq:dE1}
\end{equation}
where $c_{mn}=c_{nm}$ and $u_m = (u_{ij}+u_{ji})/2$. For Coulomb crystals with cubic symmetry such as BCC and FCC lattices, we have $c_{11}=c_{22}=c_{33}$, $c_{12}=c_{21}=c_{13}=c_{31}=c_{23}=c_{32}$, and $c_{44}=c_{55}=c_{66}$.

To estimate $c_{11}-c_{12}$, we carry out the volume-preserving deformations similar as in Refs.~\cite{Ogata1990_PRA42-4867, Caplan2018_PRL121-132701} along axis $i$ ($=x$, $y$, $z$), i.e.,
\begin{equation}
D_1:\ \ \ u_{ii} = \left( 1-\frac{\varepsilon}{2} \right) ^{-2}-1, \ \  \left.u_{jj}\right|_{j\neq i} = -\frac{\varepsilon}{2}. \label{eq:def_drop}
\end{equation}
The BCC lattice ($\varepsilon=0$) can evolve into FCC lattice ($\varepsilon=\varepsilon_\mathrm{F}=2-2^{5/6}\approx 0.2182$) by increasing $\varepsilon$, i.e., the Bain path~\cite{Bain1924_TAIMME70-25}. The elastic constants can then be fixed with
\begin{equation}
c_{11}-c_{12}=
 \left\{\begin{array}{l}
   \left. \frac{2}{3}\frac{\mbox{d}^2\delta E}{\mbox{d}\varepsilon^2} \right|_{\varepsilon=0}\ (\mathrm{BCC}); \\
   \left. \frac{2^{2/3}}{3}\frac{\mbox{d}^2\delta E}{\mbox{d}\varepsilon^2} \right|_{\varepsilon=\varepsilon_\mathrm{F}}\ (\mathrm{FCC}). \\
 \end{array}\right. \label{eq:c11mc12}
\end{equation}
To estimate $c_{44}$ of BCC and FCC lattices, we apply the deformation~\cite{Ogata1990_PRA42-4867}
\begin{equation}
D_2:\ \ \ u_{xy} = u_{yx} = \frac{\varepsilon}{2}, \ \  u_{zz} = \frac{4}{4-\varepsilon^2}-1,  \label{eq:def_drop_shear0}
\end{equation}
which is the same as the following deformation by rotating the lattices $45^\circ$ along $z$-axis, i.e.,
\begin{equation}
D_3:\ \ \ u_{xx} = -u_{yy} = \frac{\varepsilon}{2}, \ \  u_{zz} = \frac{4}{4-\varepsilon^2}-1.  \label{eq:def_drop_shear}
\end{equation}
Then to the leading order Eq.~(\ref{eq:dE1}) is reduced into $\delta E=c_{44}\varepsilon^2/2$, we thus have
\begin{equation}
  c_{44}= 2\delta E(\varepsilon)/\varepsilon^2, \label{eq:c44}
\end{equation}
which is valid for small enough $\varepsilon$.

\section{\label{sec:res}Results and discussions}

\begin{figure}
\includegraphics[width=\linewidth]{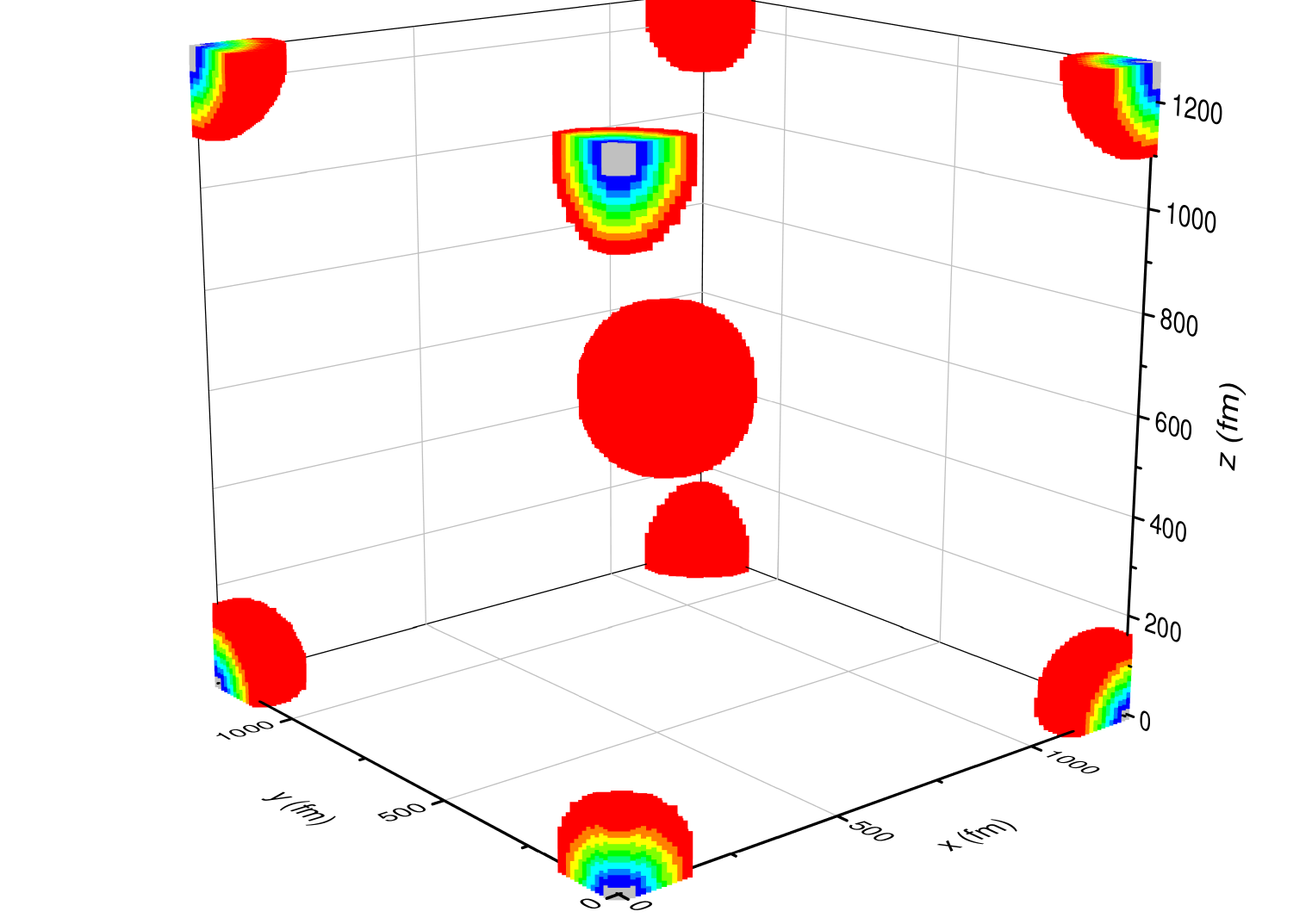}
\caption{\label{Fig:rhop} Proton density profile for a BCC lattice at an ion ($^{12}$C) number density $n_d=10^{-9}$ fm$^{-3}$ and width of Gaussian wave function $b=100$ fm for Eq.~(\ref{eq:Gaus}).}
\end{figure}

To obtain the nonuniform structures of Coulomb crystals, we first fix the position of nucleus $i$  according to the lattice structures. As an example, in Fig.~\ref{Fig:rhop} we present the proton density profile of $^{12}$C forming a BCC lattice inside a unit cell, which are fixed by taking $n_d=10^{-9}$ fm$^{-3}$ and $b=100$ fm.

\begin{figure}
\includegraphics[width=\linewidth]{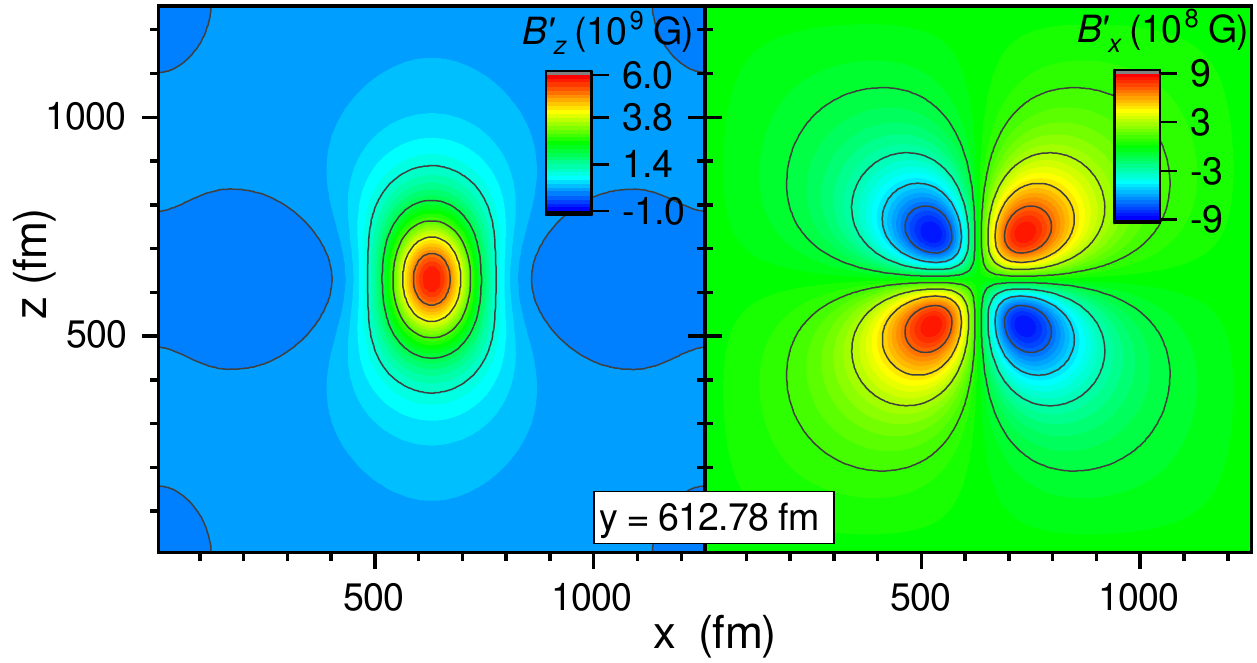}
\caption{\label{Fig:B2D} Induced magnetic fields $B_z'= B_z - B_{z0}$ and $B_x'= B_x - B_{x0}$ on the $x$-$z$ plane ($y= 612.78$ fm) of the unit cell illustrated in Fig.~\ref{Fig:rhop}, which arise due to the magnetization of electrons ($\bm{m}_i=0$ for $^{12}$C) under an external magnetic field $B_{z0} = 10^{15}$ G.}
\end{figure}

The density distributions of electrons are then obtained by fulfilling the constancy of chemical potentials, i.e.,
\begin{equation}
\mu_e(\bm{r}) =E^f_e - e A_0 = \rm{constant}, \label{eq:chem}
\end{equation}
where the electromagnetic fields are fixed by solving Eqs.~(\ref{eq:KG_A0}) and (\ref{eq:KG_mag}). Based on the Fermi energy $E^f_e(\bm{r})$ obtained with Eq.~(\ref{eq:chem}), the density distribution $\rho_e(\bm{r})$ is then determined by Eq.~(\ref{eq:rhoi_c}). In practice, Eqs.~(\ref{eq:KG_A0}) and (\ref{eq:KG_mag}) are solved iteratively inside a 3D periodic cell with discretized space coordinate, i.e.,
\begin{enumerate}
  \item \label{item:itr_1} Assume initial density distributions of electrons at a given total electron number;
  \item \label{item:itr_2} Solve Eqs.~(\ref{eq:KG_A0}) and (\ref{eq:KG_mag}) with relexation Method;
  \item \label{item:itr_3} Prepare the density distributions of fermions with the imaginary time step method~\cite{Levit1984_PLB139-147};
  \item \label{item:itr_4} Go to step~\ref{item:itr_2} until convergence is reached;
  \item \label{item:itr_5} Obtain the energy using Eq.~(\ref{eq:energy}).
\end{enumerate}
In the current study the cell is divided into $110^3$ grid points adopting a fixed grid width, where the numerical convergency is checked by varying the number of grid points. Adopting an external magnetic field $B_{z0} = 10^{15}$ G, the density profile of electrons is then fixed via iteration, which leads to nonzero magnetization according to Eq.~(\ref{eq:Mi_c}). Due to the nonuniform density distribution of electrons, the magnetization becomes nonuniform and hence there exist nonuniform induced magnetic fields inside the unit cell. In Fig.~\ref{Fig:B2D} we present the obtained induced magnetic field corresponding to the unit cell indicated in Fig.~\ref{Fig:rhop}, which is fixed by Eq.~(\ref{eq:Bcal}). Such a nonuniform magnetic field is expected to alter the properties of Coulomb crystals according to Eq.~(\ref{Eq:hydroeq}).

\subsection{Width of Gaussian wave function for nuclei}

\begin{figure*}
\begin{minipage}[t]{0.51\linewidth}
\centering
\includegraphics[width=\textwidth]{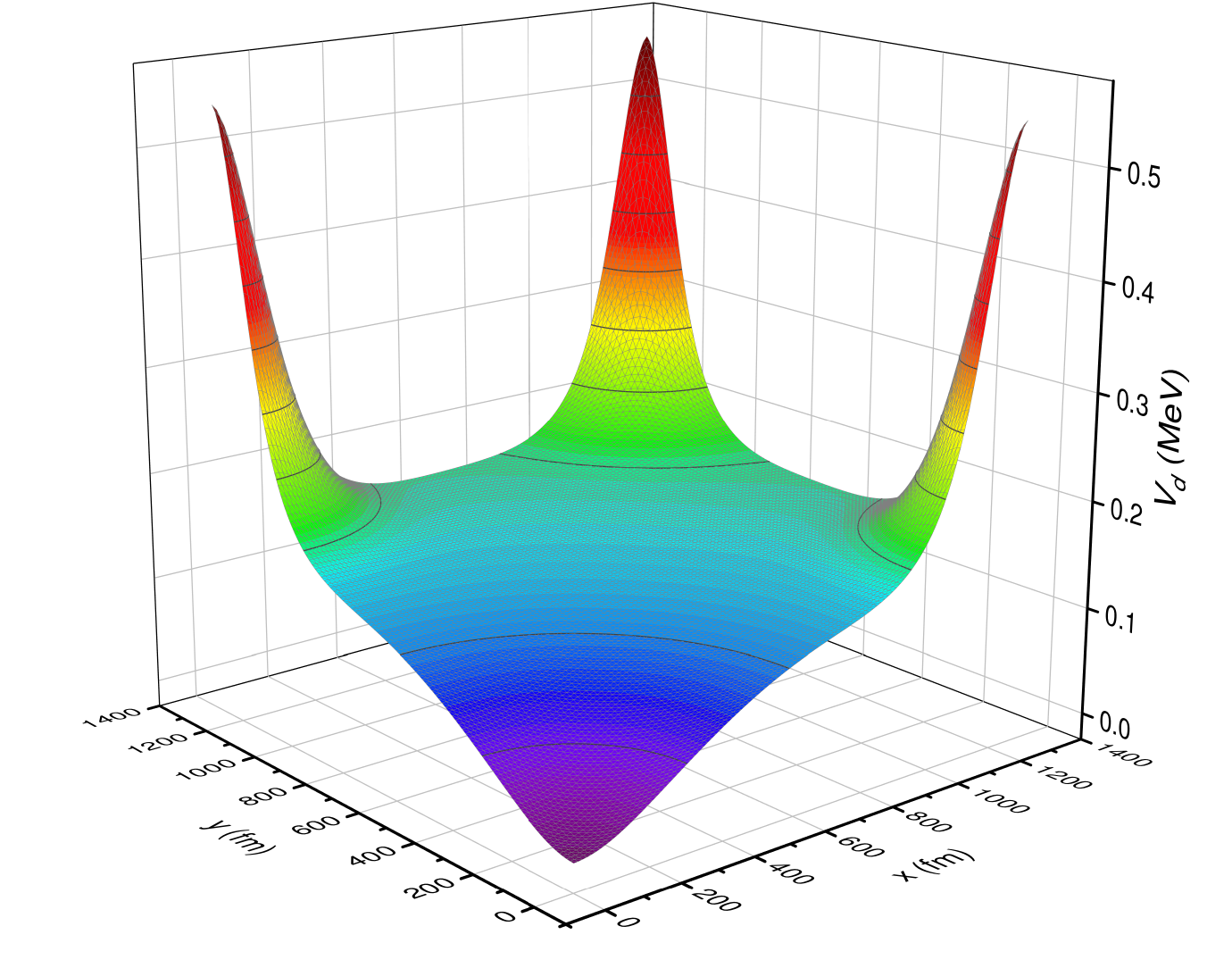}
\end{minipage}%
\hfill
\begin{minipage}[t]{0.48\linewidth}
\centering
\includegraphics[width=\textwidth]{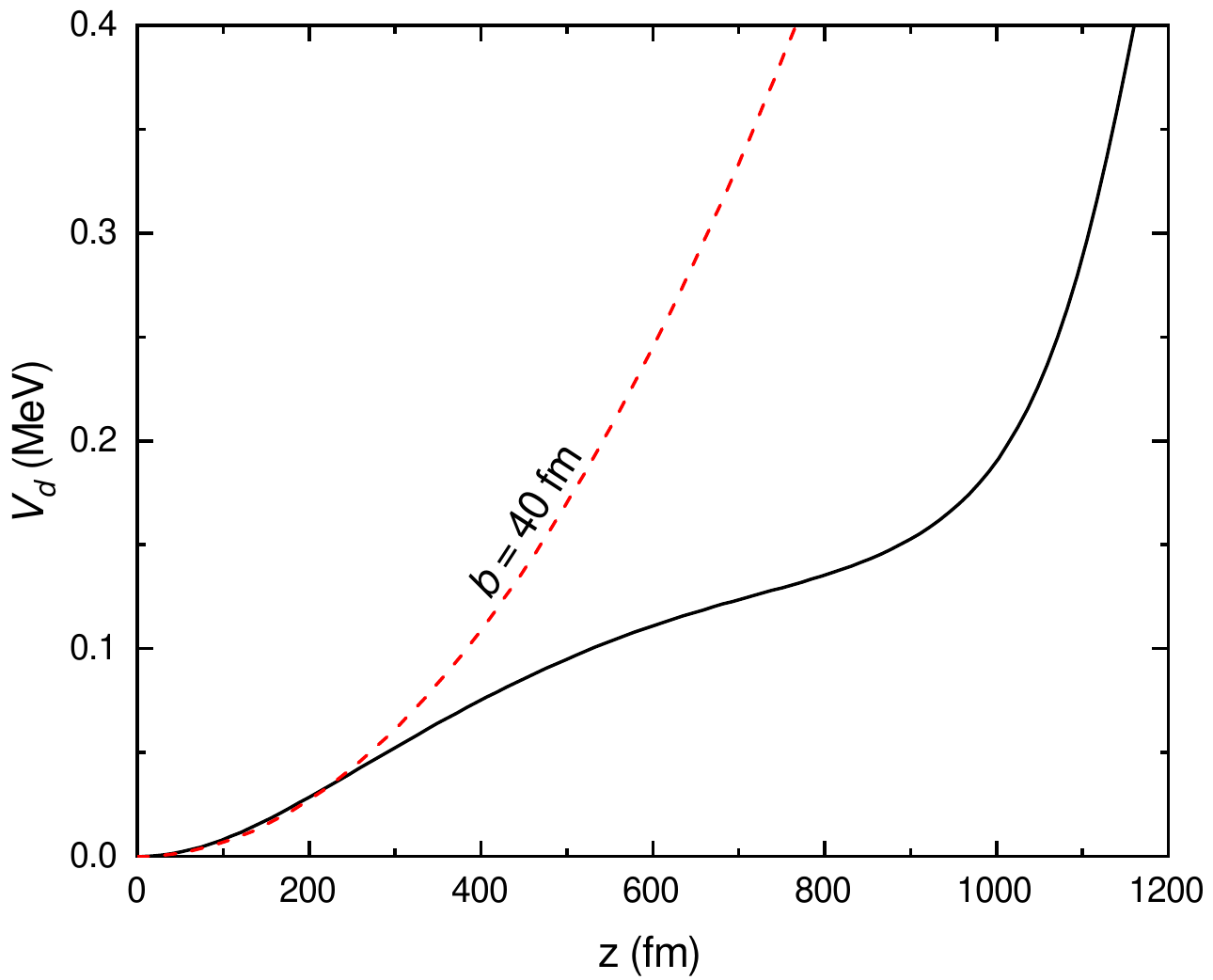}
\end{minipage}
\caption{\label{Fig:PES} Potential energies for a nucleus ($^{12}$C) located at $x=y=z=0$ on the $x$-$y$ plane ($z=0$, left panel) and along the $z$ axis ($x=y=0$, right panel), which correspond to a BCC lattice illustrated in Fig.~\ref{Fig:rhop} with an external magnetic field $B_{z0} = 10^{15}$ G. The red dashed curve in the right panel indicates the potential of a harmonic oscillator determined by Eq.~(\ref{eq:VHO}) with $b = 40$ fm, which well reproduces the potential at $r\lesssim 230$ fm.}
\end{figure*}

We first examine the impact of width of the gaussian wave packet for nuclei. By taking $b=100$ fm for Eq.~(\ref{eq:Gaus}) and assuming a BCC lattice with ion ($^{12}$C) number density $n_d=10^{-9}$ fm$^{-3}$  subjected to an external magnetic field $B_{z0} = 10^{15}$ G, the potential of a nucleus located at $x=y=z=0$ can be fixed by
\begin{equation}
V_d(\bm{r}) =q_i (A_0-A_{0,i}) - \bm{m}_i\cdot (\bm{B} - \bm{B}_i), \label{eq:Vd}
\end{equation}
where $A_{0,i}$ and $\bm{B}_i$ are the Coulomb potential and magnetic field originated from nucleus $i$ located at $x=y=z=0$. The obtained results is then presented in Fig.~\ref{Fig:PES}. Evidently, the potential of a nucleus inside the Coulomb crystal can be well approximated by the potential of a harmonic oscillator in Eq.~(\ref{eq:VHO}), while at $r\gtrsim 230$ fm the deviation grows due to the screening effects of electrons as $n_e(r)$ decreases with $r$. Note that we have used a slightly larger value with $b = 100$ fm for the gaussian wave packet, which is less numerical demanding and may be the case for nonzero temperatures.

\begin{figure}
\includegraphics[width=\linewidth]{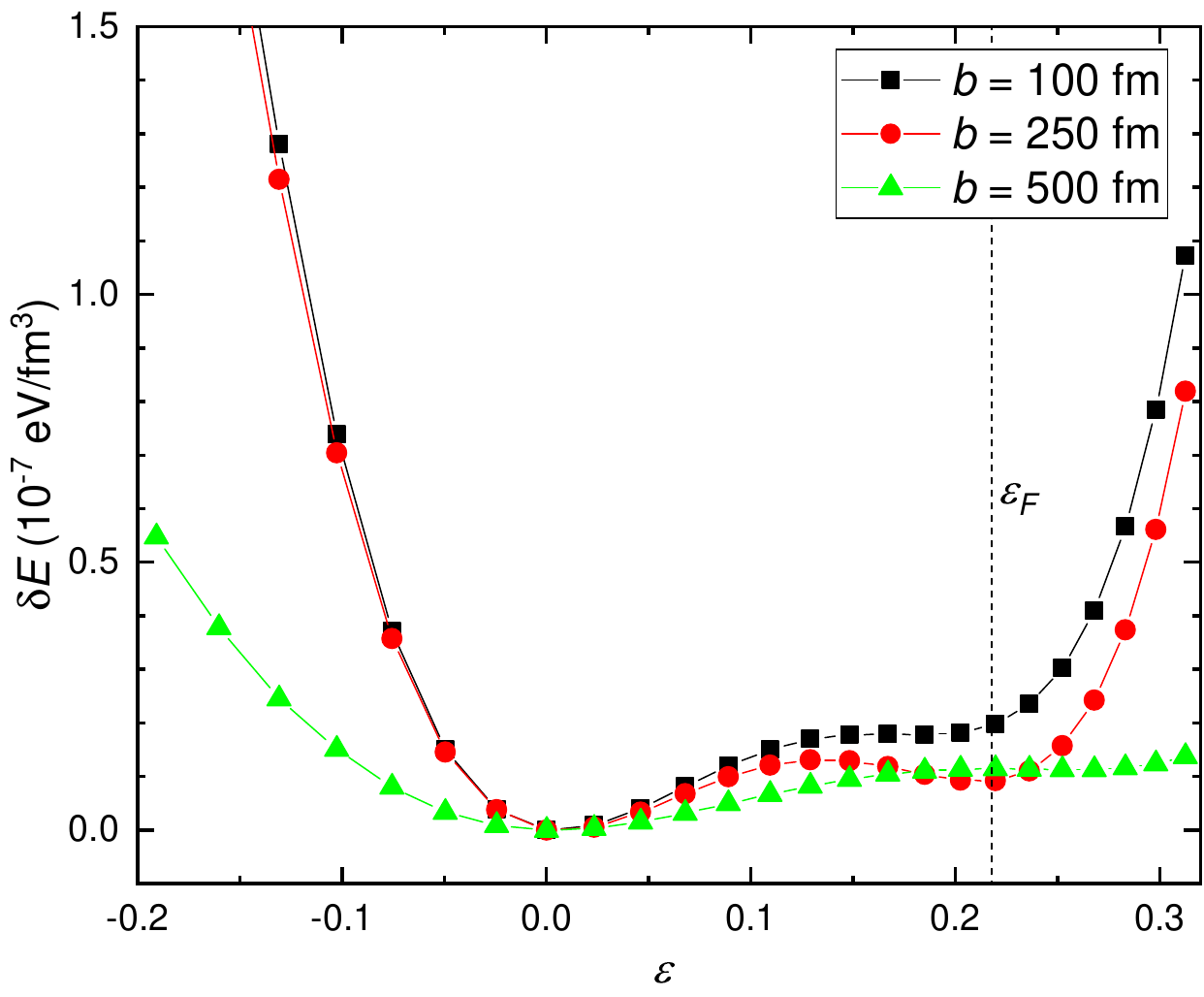}
\caption{\label{Fig:DEBtoF_b} Variation of energy density as a function of $\varepsilon$ employing deformation $D_1$, which are obtained adopting $b=100$, 250, and 500 fm for Eq.~(\ref{eq:Gaus}) at an ion ($^{12}$C) number density $n_d=10^{-9}$ fm$^{-3}$ under external magnetic field $B_{z0} = 10^{15}$ G. As $\varepsilon$ increases, the BCC lattice ($\varepsilon=0$) evolves into FCC lattice ($\varepsilon=\varepsilon_\mathrm{F}$).}
\end{figure}

In Fig.~\ref{Fig:DEBtoF_b} we present the variation of energy density as a function of $\varepsilon$ employing deformation $D_1$ along $z$-axis, where different widths for the Gaussian wave function in Eq.~(\ref{eq:Gaus}) are employed, i.e., $b=100$, 250, and 500 fm. Same as Fig.~\ref{Fig:PES}, an ion ($^{12}$C) number density $n_d=10^{-9}$ fm$^{-3}$ under external magnetic field $B_{z0} = 10^{15}$ G is employed. Evidently, the BCC lattice becomes softer as $b$ increases, where the elastic constant decreases. In particular, according to Eq.~(\ref{eq:c11mc12}), the corresponding elastic constant is $c_{11}-c_{12}=0.06475\mu_0$, $0.06052\mu_0$, and $0.01669\mu_0$ at $b=100$, 250, and 500 fm, respectively. The increment of the width of Gaussian wave function for nuclei could be attributed to zero-point ion vibrations and finite temperatures, which effectively reduces the elastic constants of Coulomb crystals~\cite{Ogata1990_PRA42-4867, Baiko2011_MNRAS416-22}. Note that even at the melting temperature with $\Gamma_m\approx175$ the electron gas is still highly degenerate due to the large Fermi temperature~\cite{Fantina2020_AA633-A149}, which justifies the assumption of zero temperature for electrons.

\subsection{Madelung constant}
As indicated in Eq.~(\ref{eq:Et}), the static-lattice binding energy density of a Coulomb crystal is
\begin{equation}
  E_0 = K_M \mu_{0}, \label{eq:E0}
\end{equation}
where the Madelung constant $K_M$ was extensively investigated and well understood~\cite{Coldwell-Horsfall1960_JMP1-395, Brush1966_JCP45-2102, Baiko2001_PRE64-57402}. For point nuclei embedded in an uniform electron gas, the Madelung constant reads~\cite{Baiko2001_PRE64-57402}
\begin{equation}
K_M=
 \left\{\begin{array}{l}
   -0.895929255682\ (\mathrm{BCC}); \\
   -0.895873615195\ (\mathrm{FCC}). \\
 \end{array}\right. \label{eq:KM}
\end{equation}
Due to the presence of Coulomb potential, the electron gas becomes nonuniform and the polarization effects should be included, which was extensively investigated and treated as a correction $\sigma(Z)$ indicated in Eq.~(\ref{eq:plrz})~\cite{Chamel2016_PRD93_63001}. For $^{12}$C, Eq.~(\ref{eq:plrz}) gives $\sigma(6)=1.003997821415$, which is negligible in our current study.

\begin{figure}
\includegraphics[width=\linewidth]{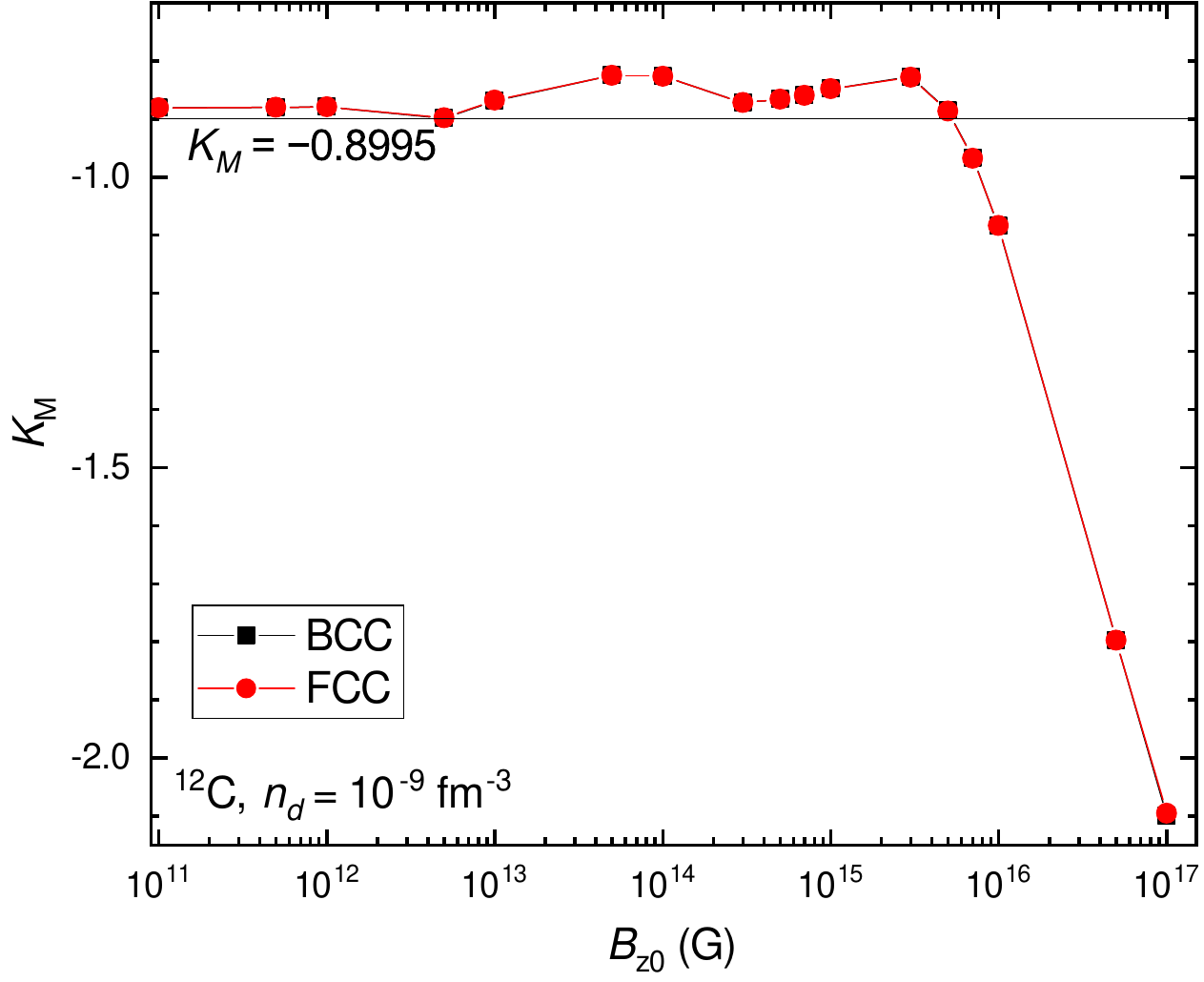}
\caption{\label{Fig:Madelung} Madelung constants for BCC and FCC lattices as functions of external magnetic field, which are obtained employing $b=100$ fm and $n_d=10^{-9}$ fm$^{-3}$. The solid horizontal line indicates the value obtained using Eq.~(\ref{eq:KM}) with polarization correction.}
\end{figure}

\begin{table}
\caption{\label{table:Madelung} Madelung constants (with polarization effects) for BCC and FCC lattices as indicated in Fig.~\ref{Fig:Madelung}, which are obtained at an ion ($^{12}$C) number density $n_d=10^{-9}$ fm$^{-3}$ employing $b=100$ fm.}
\begin{tabular}{c|cc|c|cccc} \hline \hline
          $B_{z0}$ & $K_M^\mathrm{BCC}$&   $K_M^\mathrm{FCC}$  &          $B_{z0}$ & $K_M^\mathrm{BCC}$&   $K_M^\mathrm{FCC}$  \\
            G      &     $\mu_0$       &   $\mu_0$             &            G      &     $\mu_0$       &   $\mu_0$             \\ \hline
 $1\times 10^{11}$ &  $-0.88035$       & $-0.88022$            & $7\times 10^{14}$ &  $-0.85892$       & $-0.85882$            \\
 $5\times 10^{11}$ &  $-0.88004$       & $-0.87992$            & $1\times 10^{15}$ &  $-0.84808$       & $-0.84798$            \\
 $1\times 10^{12}$ &  $-0.87872$       & $-0.87848$            & $3\times 10^{15}$ &  $-0.82753$       & $-0.82822$            \\
 $5\times 10^{12}$ &  $-0.89843$       & $-0.89792$            & $5\times 10^{15}$ &  $-0.88533$       & $-0.88639$            \\
 $1\times 10^{13}$ &  $-0.86784$       & $-0.86776$            & $7\times 10^{15}$ &  $-0.96736$       & $-0.96728$            \\
 $5\times 10^{13}$ &  $-0.82491$       & $-0.82460$            & $1\times 10^{16}$ &  $-1.08389$       & $-1.08363$            \\
 $1\times 10^{14}$ &  $-0.82644$       & $-0.82614$            & $5\times 10^{16}$ &  $-1.79683$       & $-1.79726$            \\
 $3\times 10^{14}$ &  $-0.87131$       & $-0.87116$            & $1\times 10^{17}$ &  $-2.10034$       & $-2.09462$            \\
 $5\times 10^{14}$ &  $-0.86586$       & $-0.86572$            &   &         &             \\
\hline
\end{tabular}
\end{table}

By comparing the energy densities of Coulomb crystals and the corresponding uniform matter, the binding energy density can be estimated. The Madelung constants (with polarization effects) can then be fixed according to Eq.~(\ref{eq:E0}), where in Fig.~\ref{Fig:Madelung} we present the obtained results for BCC and FCC lattices as functions of external magnetic field $B_{z0}$. At small $B_{z0}$ ($\lesssim 3\times 10^{14}$ G), $K_M$ fluctuates with $B_{z0}$ due to Landau quantization and is slightly larger than that obtained in uniform electron background~\cite{Baiko2001_PRE64-57402}. As the external magnetic field strength further increases, electrons occupy the lowest Landau level, so that $K_M$ increases until reaching its maximum value at $B_{z0}= 3\times 10^{15}$ G and then declines at larger $B_{z0}$, i.e., the binding energy decreases with $B_{z0}$ at $B_{z0}\lesssim 3\times 10^{15}$ G and increases at larger external magnetic fields. Same as nonmagnetic scenarios indicated in Eq.~(\ref{eq:KM}), the Madelung constants for BCC and FCC lattices are generally the same, while that of BCC is slightly smaller than FCC at $B_{z0} < 3\times 10^{15}$ G. At larger $B_{z0}$ ($\geq 3\times 10^{15}$ G), nevertheless, the FCC lattice may be more stable.

\subsection{Elastic properties}

The elastic properties of magnetized Coulomb crystals can be examined by carrying out deformations as illustrated in Sec.~\ref{sec:the_elestic}. Adopting deformation $D_1$ along $z$ axis with Eq.~(\ref{eq:def_drop}), the elastic constants $c_{11}-c_{12}$ for BCC and FCC latices can be fixed using Eq.~(\ref{eq:c11mc12}), while that of $c_{44}$ can be determined by Eq.~(\ref{eq:c44}) employing deformation $D_2$ with Eq.~(\ref{eq:def_drop_shear0}).

\begin{figure}
\includegraphics[width=\linewidth]{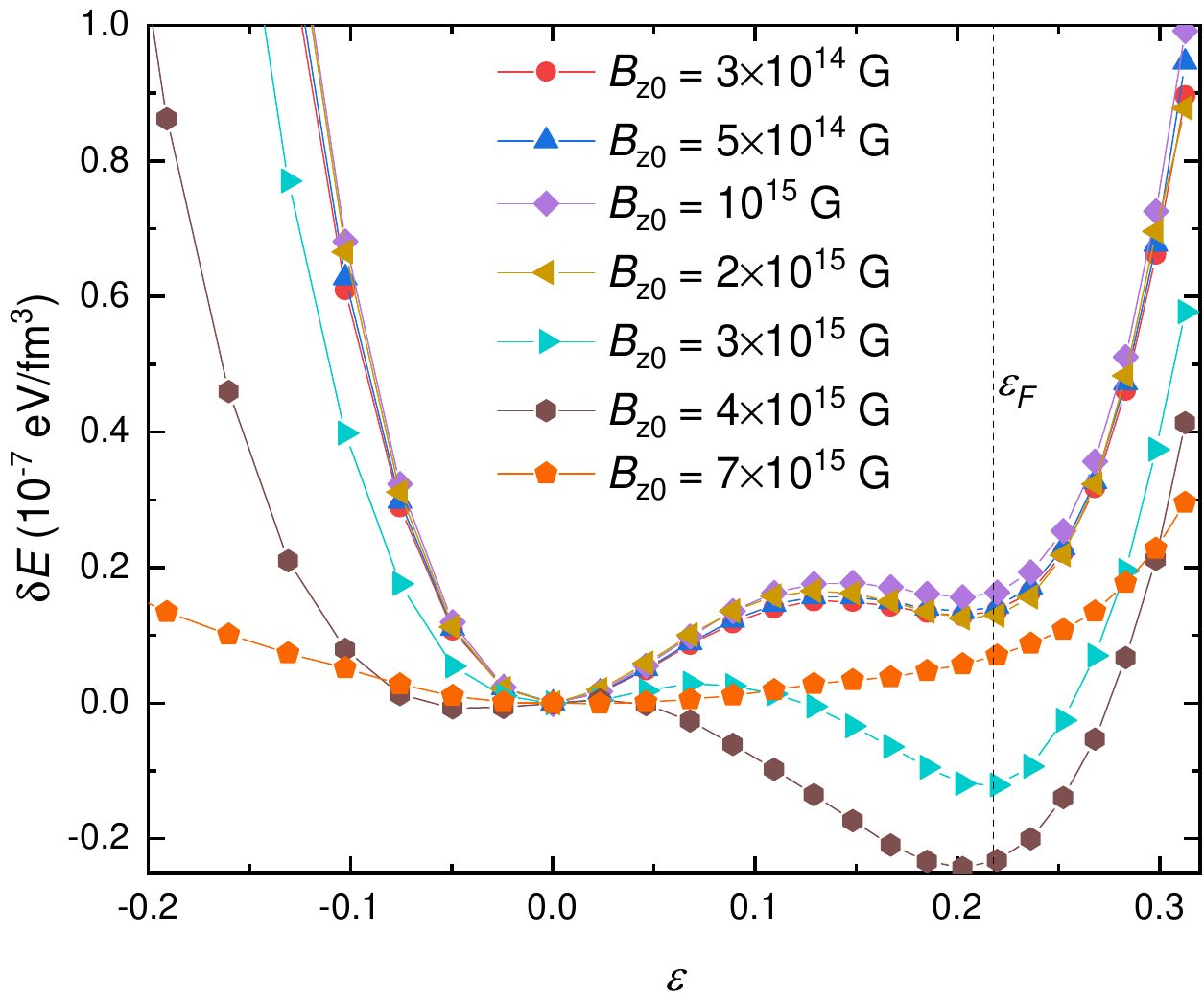}
\caption{\label{Fig:DE_BZ} Same as Fig.~\ref{Fig:DEBtoF_b} but adopting $b=100$ fm and various values for $B_{z0}$.}
\end{figure}

In Fig.~\ref{Fig:DE_BZ} we present the variation of energy density as a function of $\varepsilon$ employing deformation $D_1$ along $z$-axis, where different external magnetic fields $B_{z0}$ are employed. The elastic properties varies little with $B_{z0}$ at $B_{z0}< 3\times 10^{15}$ G since the variations of energy density as functions of $\varepsilon$ generally coincide with each other. At larger $B_{z0}$, the elastic properties change quickly with much larger variations in $\delta E(\varepsilon)$. As $\varepsilon$ increases, the BCC lattice ($\varepsilon=0$) evolves into FCC lattice ($\varepsilon=\varepsilon_\mathrm{F}\approx 0.2182$). It is found that the BCC lattices are generally more stable than that of FCC lattices at $B_{z0}< 3\times 10^{15}$ G, where the situation is reversed at larger $B_{z0}$, in coincidence with those listed in Table~\ref{table:Madelung}. At $B_{z0}\gtrsim 7\times 10^{15}$ G, the FCC lattice is no longer (meta)stable as there does not exist a local minimum at $\varepsilon=\varepsilon_\mathrm{F}$, while the BCC lattice at $\varepsilon=0$ becomes softer. If we further increase $B_{z0}$, $\delta E(\varepsilon)$ even becomes multi-valued and it is difficult to find a stable lattice structure, while in Fig.~\ref{Fig:DE_BZ} only one branch is indicated.

\begin{figure}
\includegraphics[width=\linewidth]{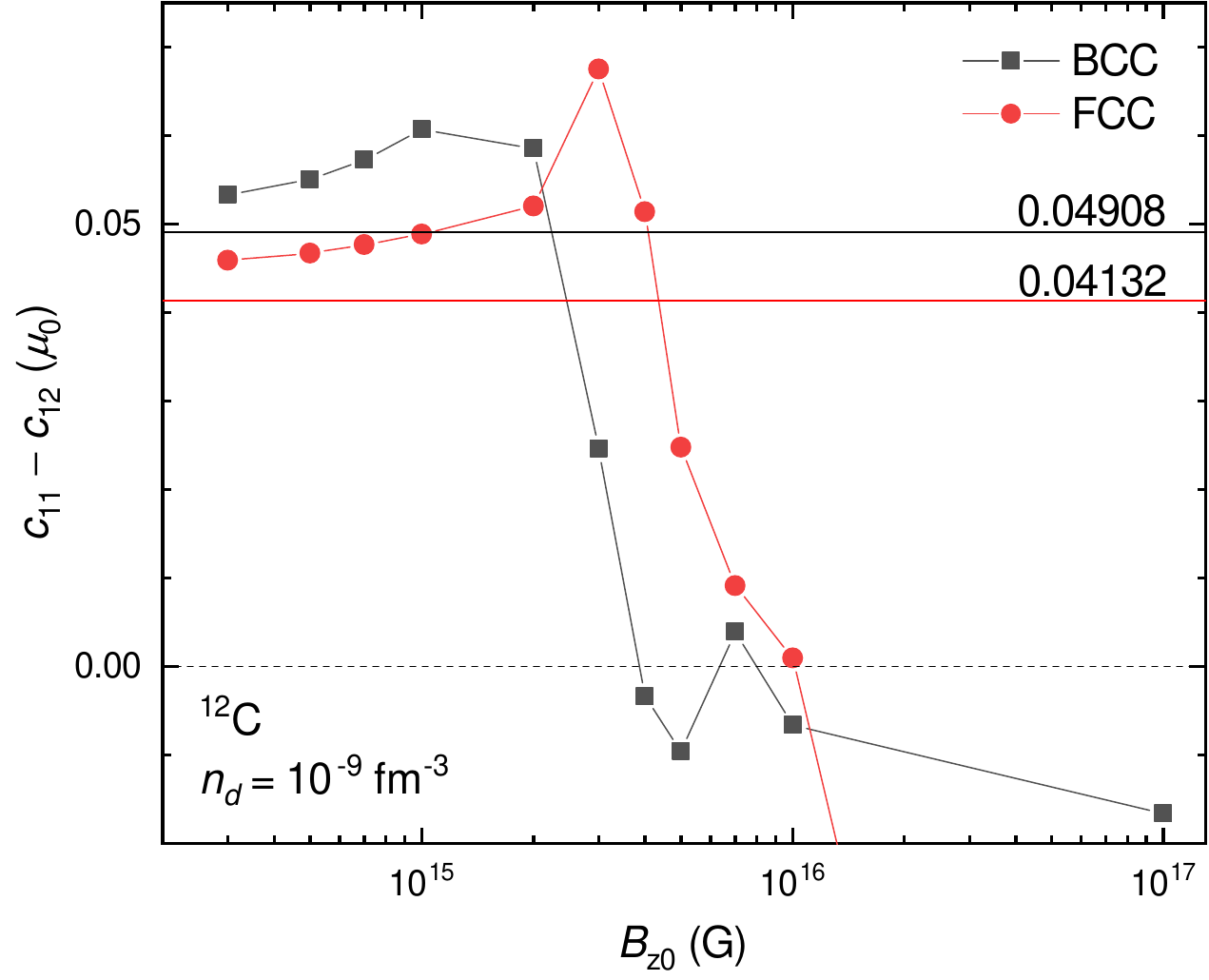}
\caption{\label{Fig:c11mc12} Elastic constants $c_{11}-c_{12}$ for BCC and FCC lattices, which are obtained adopting $b=100$ fm at an ion ($^{12}$C) number density $n_d=10^{-9}$ fm$^{-3}$ under various external magnetic fields.}
\end{figure}

\begin{table}
\caption{\label{table:Elas} Elastic constants for BCC and FCC lattices as indicated in Fig.~\ref{Fig:DE_BZ}, which are obtained at an ion ($^{12}$C) number density $n_d=10^{-9}$ fm$^{-3}$ employing $b=100$ fm. }
\begin{tabular}{c|rr|rrccc} \hline \hline
                  &  $\mathrm{BCC}$  &   $\mathrm{FCC}$  &  $\mathrm{BCC}$  &   $\mathrm{FCC}$    \\ \hline
 $B_{z0}$      &  \multicolumn{2}{c|}{$c_{11}-c_{12}$}   &  \multicolumn{2}{c}{$c_{44}$}    \\
          G       &     $\mu_0$      &   $\mu_0$         &    $\mu_0$    &   $\mu_0$         \\ \hline
$3\times 10^{14}$ &  $0.05331$       & $0.04589$         &    0.18461    &   $0.18840$       \\
$5\times 10^{14}$ &  $0.05505$       & $0.04674$         &    0.18955    &   $0.19584$       \\
$7\times 10^{14}$ &  $0.05726$       & $0.04767$         &    0.19483    &   $0.20162$       \\
$1\times 10^{15}$ &  $0.06067$       & $0.04886$         &    0.20298    &   $0.21047$       \\
$2\times 10^{15}$ &  $0.05853$       & $0.05205$         &    0.20665    &   $0.21506$       \\
$3\times 10^{15}$ &  $0.02461$       & $0.06752$         &    0.16383    &   $0.15917$       \\
$4\times 10^{15}$ &  $-0.00336$      & $0.05137$         &    0.09867    &   $0.08907$       \\
$5\times 10^{15}$ &  $-0.00957$      & $0.02478$         &    0.04508    &   $0.03562$       \\
$7\times 10^{15}$ &  $0.00391$       & $0.00915$         &    0.00190    &   $-0.00054$      \\
$1\times 10^{16}$ &  $-0.00662$      & $0.00097$         &    0.00550    &   $0.00648$       \\
$1\times 10^{17}$ &  $-0.01662$      & $-0.17520$        &    $-0.39423$ &   $0.00220$       \\
\hline
\end{tabular}
\end{table}

Based on the results presented in Fig.~\ref{Fig:DE_BZ}, the elastic constants $c_{11}-c_{12}$ for BCC and FCC lattices can then be estimated with Eq.~(\ref{eq:c11mc12}), where in Fig.~\ref{Fig:c11mc12} and Table~\ref{table:Elas} we present the obtained values at various external magnetic fields $B_{z0}$. At $B_{z0} \leq 10^{15}$ G, the elastic constants $c_{11}-c_{12}$ increases with $B_{z0}$, while at $10^{15}\ \mathrm{G} \lesssim B_{z0} \lesssim 3\times 10^{15}$ G the elastic constants $c_{11}-c_{12}$ for BCC lattices decrease as the FCC lattices become more stable. At $B_{z0} > 3\times 10^{15}$ G the elastic constants $c_{11}-c_{12}$ decrease with $B_{z0}$ and even become negative at $B_{z0} \gtrsim 10^{16}$ G, suggesting that the corresponding lattice structures are no longer stable. Note that as $\delta E(\varepsilon)$ becomes multi-valued at $B_{z0}\gtrsim 7\times 10^{15}$ G in Fig.~\ref{Fig:DE_BZ}, the elastic constants are estimated by randomly pick a branch for $\delta E(\varepsilon)$, which could vary slightly if we adopt other branches for $\delta E(\varepsilon)$. In particular, there exists a local peak for the elastic constants $c_{11}-c_{12}$ of BCC lattices at $B_{z0} = 7\times 10^{15}$ G, which is small and may become negative due to numerical uncertainties.

\begin{figure}
\includegraphics[width=\linewidth]{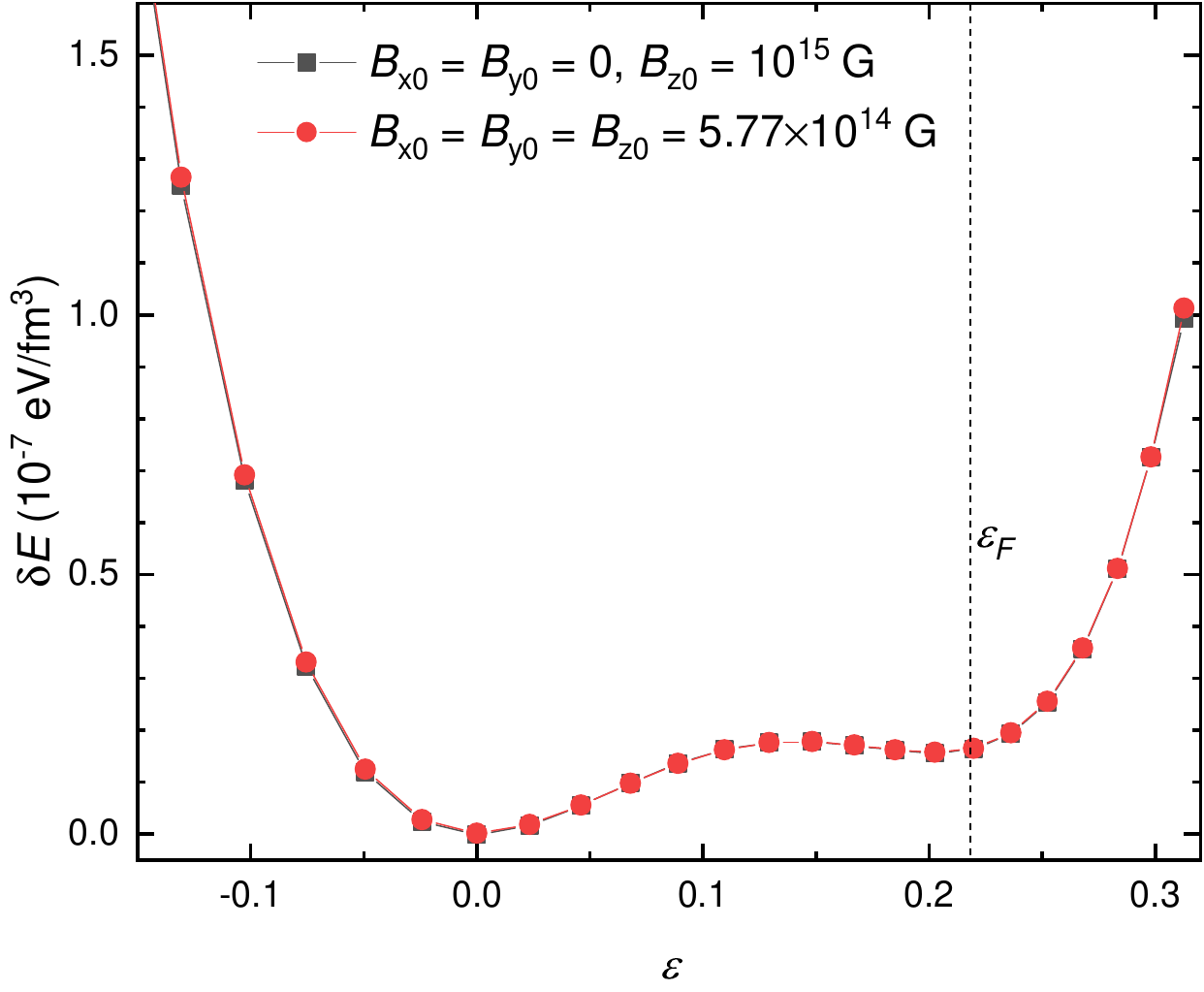}
\caption{\label{Fig:DE_BZBxy} Same as Fig.~\ref{Fig:DE_BZ} but adopting external magnetic fields $B_{0}=10^{15}$ G with differently orientations.}
\end{figure}

The investigation on the elastic properties of Coulomb crystals in Fig.~\ref{Fig:DE_BZ} are carried out under external magnetic fields at fixed direction along $z$-axis, which is expected to break the cubic symmetry. To examine if the cubic symmetry is still preserved, we further investigate the elastic properties of Coulomb crystals with different external magnetic field orientations, where in Fig.~\ref{Fig:DE_BZBxy} we present the obtained results with $B_{0}=10^{15}$ G but oriented differently. Evidently, the variations of energy density $\delta E(\varepsilon)$ under the two circumstances are almost identical, suggesting that the elastic properties are barely altered by the orientation of external magnetic fields.

\begin{figure}
\includegraphics[width=\linewidth]{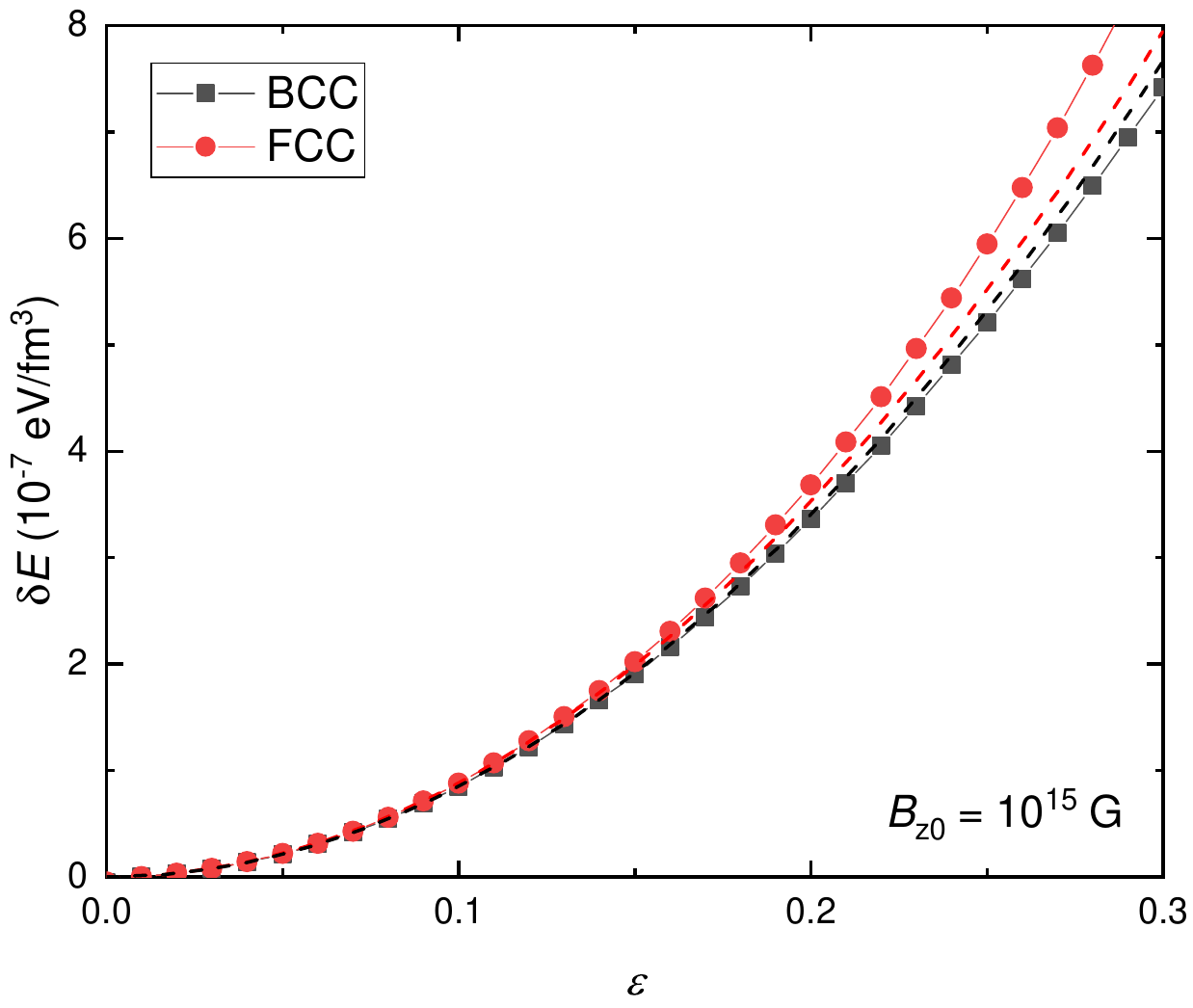}
\caption{\label{Fig:DE_c44} Variation of energy density as a function of $\varepsilon$ employing deformation $D_2$, which are obtained adopting $b=100$ fm at an ion ($^{12}$C) number density $n_d=10^{-9}$ fm$^{-3}$ under external magnetic field $B_{z0} = 10^{15}$ G. The dashed curves indicate the functional form $\delta E=c_{44}\varepsilon^2/2$ from elastic theory, where the elastic constant $c_{44}$ is indicated in Table~\ref{table:Elas}.}
\end{figure}

Next we examine the elastic properties employing deformation $D_2$, where the elastic constant $c_{44}$ can be estimated with Eq.~(\ref{eq:c44}). In Fig.~\ref{Fig:DE_c44} we present the variation of energy density as a function of $\varepsilon$ employing deformation $D_2$, which are compared with the functional form $\delta E=c_{44}\varepsilon^2/2$ from elastic theory. Evidently, at $\varepsilon\lesssim 0.18$ a constant elastic constant $c_{44}$ well describes the variation of energy density under deformation $D_2$.

\begin{figure}
\includegraphics[width=\linewidth]{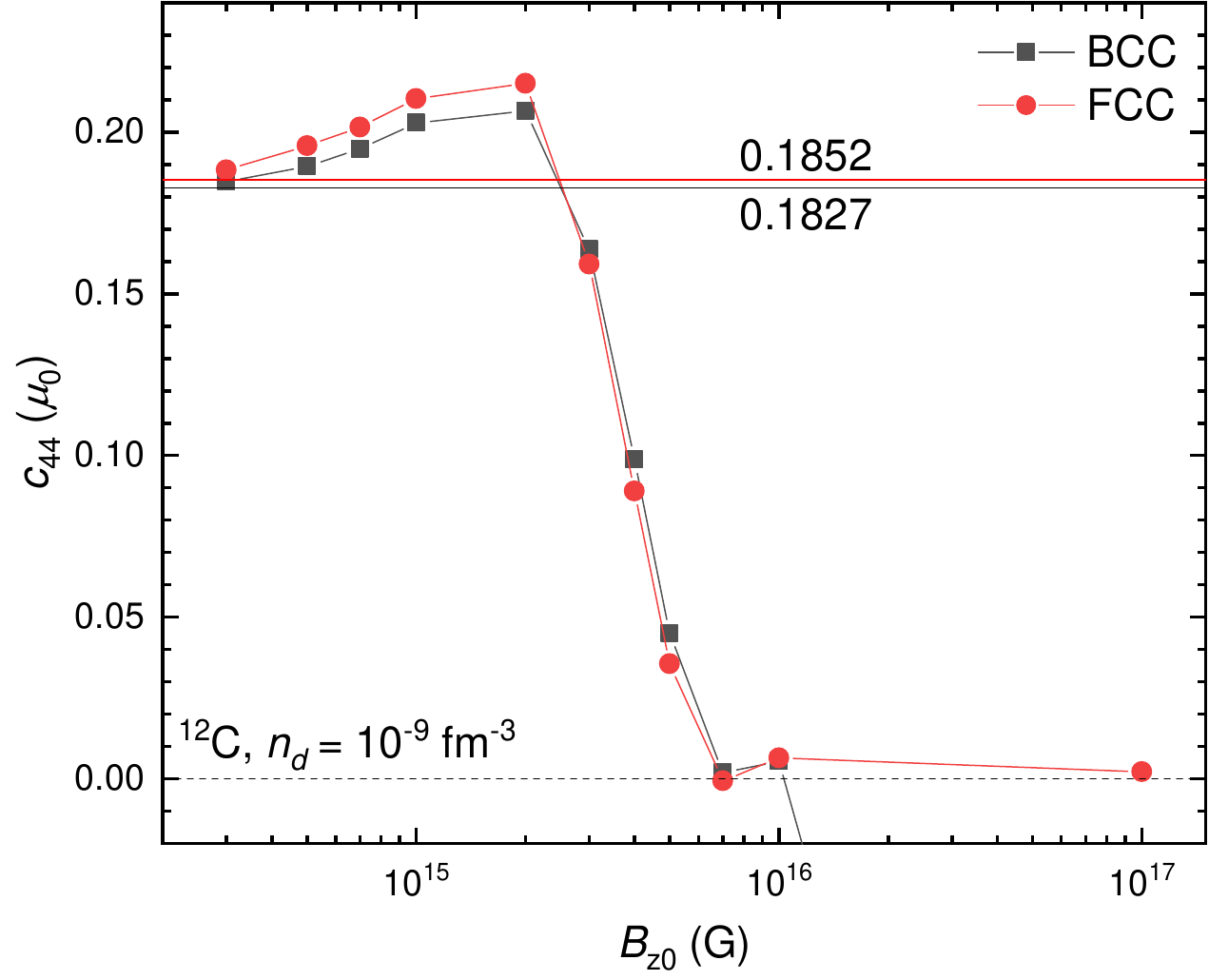}
\caption{\label{Fig:c44} Elastic constants $c_{44}$ for BCC and FCC lattices, which are obtained adopting $b=100$ fm at an ion ($^{12}$C) number density $n_d=10^{-9}$ fm$^{-3}$ under various external magnetic fields.}
\end{figure}

We then present the obtained elastic constants $c_{44}$ for BCC and FCC lattices in Table~\ref{table:Elas} and Fig.~\ref{Fig:c44}. It is found that the elastic constants $c_{44}$ generally increase with $B_{z0}$ until reaching their peaks at $B_{z0} = 2\times 10^{15}$ G, then decrease with $B_{z0}$ and are vanishing at $B_{z0} \geq 7\times 10^{15}$ G. This is roughly consistent with the behavior of the elastic constants $c_{11}-c_{12}$ indicated in Fig.~\ref{Fig:c11mc12}. At small external magnetic fields with $B_{z0} \leq 2\times 10^{15}$ G, under deformation $D_2$, the FCC lattice is slightly stiffer than BCC lattice with larger $c_{44}$, which nonetheless becomes softer at larger $B_{z0}$.

\section{\label{sec:con}Summary}

In this work, the properties of magnetized Coulomb crystals in neutron star crusts are investigated in a fully three-dimensional geometry with periodic boundary condition. The Thomas-Fermi approximation is employed to fix the electron density profiles, while the nonuniform magnetic fields are treated with equivalent magnetic charge method. In particular, we consider Coulomb crystals made of $^{12}$C with an ion number density $n_d=10^{-9}$ fm$^{-3}$ under various external magnetic fields. A gaussian wave function is assumed for the nuclei, where the width $b$ is attributed to the zero-point ion vibrations and finite temperature effects. Our findings are listed as follows, i.e.,
\begin{itemize}
  \item The Coulomb crystal becomes softer as the width of Gaussian wave function $b$ increases, in concordance with previous investigations~\cite{Ogata1990_PRA42-4867, Baiko2011_MNRAS416-22}.
  \item The static-lattice binding energy density of the Coulomb crystal (Madelung constant $K_M$, in unit of $\mu_0$) fluctuates with the external magnetic field $B_{z0}$ at $B_{z0}\leq 3\times 10^{14}$\,G. At higher field strengths, it decreases with $B_{z0}$ at $3\times 10^{14}\,\mathrm{G}\leq B_{z0}\lesssim 3\times 10^{15}$ G and increases at larger external magnetic fields. The BCC lattice is slightly more stable than that of FCC at $B_{z0} < 3\times 10^{15}$ G, while the FCC lattice may become more stable at larger $B_{z0}$.
  \item The elastic constants $c_{11}-c_{12}$ and $c_{44}$ for the Coulomb crystals under various external magnetic fields are examined and presented in Table~\ref{table:Elas}, which are increasing with $B_{z0}$ at $3\times 10^{14}\,\mathrm{G}\lesssim B_{z0}\lesssim 2\times 10^{15}$ G and decreasing to vanishing values at larger $B_{z0}$. It was shown that the orientation of external magnetic fields ($B_{z0} = 10^{15}$ G and $B_{x0} = B_{y0} = B_{z0} = 5.77\times 10^{14}$ G) has little impact on the elastic properties of Coulomb crystals. At $B_{z0}\gtrsim 10^{16}$, it is difficult to find a stable lattice structure.
\end{itemize}
The results presented here should be helpful to unveil the impact of strong magnetic field on the properties of Coulomb crystals in neutron star crusts.

\section*{ACKNOWLEDGMENTS}
This work was supported by the National Natural Science Foundation of China (Grant No. 12275234).

\newpage

%

\end{document}